\documentclass[manuscript]{aastex}

\usepackage{graphicx}
\usepackage{epstopdf}

\begin{document}

\title{Spectro-interferometric observations of classical nova V458 Vul 2007}

\author{Samira Rajabi\altaffilmark{1}, 
Matthew W.~Muterspaugh\altaffilmark{1,2}, Benjamin F.~Lane\altaffilmark{3}, Martin M.~Sirk\altaffilmark{4} , Stanley Browne\altaffilmark{4}, Askari Ghasempour\altaffilmark{1}, Samuel P.~Halverson\altaffilmark{5}, John G.~Kelly\altaffilmark{2}, 
Michael Williamson\altaffilmark{1}}
\altaffiltext{1}{Tennessee State University, Center of Excellence in 
Information Systems, 3500 John A. Merritt Blvd., Box No.~9501, Nashville, TN 37209-1561}
\altaffiltext{2}{Department of Mathematics and Physics, College of Arts and 
Sciences, Tennessee State University, Boswell Science Hall, Nashville, TN 37209 }
\altaffiltext{3}{Draper Laboratory, 555 Technology Square, Cambridge, MA 02139-3563}
\altaffiltext{4}{Space Sciences Laboratory, University of California, Berkeley, CA 94720 USA}
\altaffiltext{5}{Department of Astronomy and Astrophysics, The Pennsylvania State University, University Park, PA 16801}

\email{samira@coe.tsuniv.edu,matthew1@coe.tsuniv.edu}

\begin{abstract}
We used the Palomar Testbed Interferometer (PTI) to resolve 2.2 $\mu$m emission from the classical nova V458 Vul 2007 over the course of several days following its discovery on 2007 August 8.54 UT. We also obtained K-band photometric data and spectra of the nova during the early days of the outburst. We also used photometric measurements from the AAVSO database. This is a unique data set offering a 3-technique approach: high-resolution imaging, spectroscopy and photometry. Our analysis shows that the nova ejecta can be modeled as an inclined disk at low inclination i.e. low ellipticity which is consistent with the nova being in the fireball phase at which the outflowing gas is optically thick, confirmed by the presence of strong P-Cygni Balmer lines in the spectra. The expansion velocity is $\approx$1700 $\rm km\ s^{-1}$, derived from the H$\alpha$ line. By combining the nova's angular expansion rate measured by PTI with the expansion rate measured from spectroscopy, the inferred distance to the nova is 9.9-11.4 kpc. We also used the K-band fluxes and the derived size of the emission to estimate the total mass ejected from the nova $\approx 4\times 10^{-4} M_{\odot}$. The quick transition of the nova from Fe II to He/N class makes V458 Vul 2007 a hybrid nova. 
\end{abstract}

\keywords{Stars: novae, Cataclysmic variables--techniques: interferometric}

\section{Introduction}

Novae occur in close binary systems when the mass accreting from a Roche-lobe filling stellar companion onto the primary white dwarf ignites nuclear fusion in a runaway manner, giving rise to observed energetic stellar explosions \citep{Starrfield1987,Prialniketal1995,Bode2010}. The typical accretion rate through the disk is $M_{acc}=10^{-9}$ $M_{\odot}yr^{-1}$ \citep{Warner2008}. 
 Studying novae systems at different stages of their evolution improves our understanding of close binary systems. Additionally, the large amount of dust formed in different stages of  novae evolution contribute partially to the observed abundances of certain elements in the interstellar medium \citep{Gehrzetal1998}.
 One of the main questions regarding these explosions is at what distance they occur. To find this distance we need to study a nova during the early days of its explosion. Today, with the advent of Optical/IR interferometric techniques, we can achieve a very high spatial resolution of a few milliarcseconds which allows us to follow the evolution of a nova from the beginning phase of the outburst. This enables us to study the initial shaping mechanism of novae in more detail Ê\citep{Bode2002}. Some novae have already been studied using interferometers, e.g. V838 Monocerotis by PTI \citep{Lane05}, RS Ophiuchi by IOTA, Keck, PTI \citep{Monnier2006}, VLTI \citep{Chesneau2007}, PTI \citep{Lane07b} and KIN \citep{Barry2008}, V1663 Aquilae by PTI \citep{Lane07}, V1280 Scorpii by VLTI \citep{Chesneau2008} and T Pyx by CHARA/VLTI \citep{Chesneau2011} . \\
 
 Nova Vulpeculae 2007 (V458 Vul, \citep{Samus2007}) is a classical nova which was discovered on 2007 August 8.54 UT with coordinates
 $\alpha=19^{h}54^{m}24.^{s}64$, $\delta=+20^{\circ}52^{\prime}51.^{\prime\prime}9$ J2000 
 by H. Abe \citep{Nakano2007}. The light curve of V458 Vul shows a brightness of V=8 mag at the time of discovery, fading in luminosity by 2 magnitudes in 6 days after outburst \citep{Bianciardi2007,Nakamuraetal2007,Tarasova2007}, placing this nova in the very fast novae category in the Payne-Gaposchkin classification \citep{Payne1957}. One of the spectacular features of the V458 Vul light curve is  rebrightening of the nova in the early phase of its evolution \citep{Arai2009}. The nova light curve shows several peaks in the early days after its explosion as well as some quasi-periodic oscillations in the transition phase. A small fraction ($\sim 15 \%$) of novae exhibit large and small amplitude oscillations in the early phase of their evolution \citep{Retter2003}. It is not clear what causes these oscillations \citep{Warner1995,Warner2008}. \cite{Tanaka2011} studied six novae, including V458 Vul, which undergo several rebrightenings during the early phase after outburst. They confirm the re-appearance of the P-Cygni absorption features in Balmer lines and HeI lines in all brightness maximums following the first peak in the light curves of these novae. Initially the P-Cygni absorption in  Balmer line profiles at the maximum brightness is produced after the expansion of the ejected material following the nova explosion while the photosphere starts to shrink. They propose that the re-appearance of the absorption at rebrightenings  is due to the re-expansion of the photosphere and therefore enhancement of the continuum  emission rather than strengthening the line emission. Another likely scenario for explaining the oscillations in the transition phase is that these features are only seen in the light curves of novae which are intermediate polars (IPs) and are caused by the interaction of the accretion disk with the magnetosphere of the white dwarf \citep{Retter2003}. On the other hand, \cite{Shaviv2001,Shaviv2002} explains the transition phase as a natural consequence of the steady-state super-Eddington theory applied to the novae: if the luminosity is not sufficient to push the material ejected by the nova to infinity, the wind {\em stagnates}, which in turn causes the observed variable luminosities. \\ 
 
Spectroscopic studies of V458 Vul 2007 suggest that it is a hybrid nova, transitioning  from Fe II  spectral class at early days of explosion to He/N class, one month after onset of explosion, with a typical expansion velocity of $\approx$ 2000 $\rm km\ s^{-1}$ \citep{Lynch2007,Prater2007,Tarasova2007,Poggiani2008}. In September 2008, V458 Vul 2007 was detected as a highly variable supersoft X-ray source by the Swift XRT \citep{Drakeetal2008}. X-ray spectroscopy of the nova 88 days after the explosion detected hard X-ray emission which could emanate  from the shocks within the wind ejected by the nova \citep{Tsujimoto2009}.
The peculiarity of V458 Vul became even more evident when the photometric H$\alpha$ images revealed a planetary nebula around the nova central source, only the second ever nova  known to be in the center of a planetary nebula after GK Per in 1901, which was shed $\thicksim$ 14000 yrs ago, presenting V458 Vul as an interesting phenomena for studying the evolution of the planetary nebulae as well \citep{Wessonetal2008}. Following this study, \cite{Rodriguez2010} measured the orbital period to be $98.09647\pm0.00025$ min for the V458 Vul binary system. They suggest that the central binary system of the planetary nebula consists of an accreting white dwarf of mass $M_{1}\geq 1 M_{\odot}$ and a post-AGB star of mass $M_{2}\sim 0.6 M_{\odot}$ which ejected the planetary nebula. Their total estimated mass of the binary system makes V458 Vul a type Ia supernova progenitor candidate.\\

We used the Palomar Testbed Interferometer (PTI) to resolve the 2.2 $\mu m$ emission from V458 Vul on nine nights following the 2007 outburst. PTI was a NIR long baseline (85-110 m) stellar interferometer built by NASA/JPL as a testbed for developing interferometric techniques applicable to the Keck Interferometer and was located on Palomar Mountain near San Diego, California \citep{Colavitaetal1999}. It combined star light from two out of three fixed 40 cm telescopes pairwise and measured the fringe contrast or visibility. PTI achieved a very high angular resolution of $4$ mas in the K band. Therefore we were able to resolve emission at the milliarcsecond scale and place constraints on the emission size, morphology and possible ongoing mechanisms in novae explosions.\\
We used K-band fluxes along with the computed emission sizes to estimate the total ejected mass from the nova. We also used the PTI low-resolution spectrometer to estimate the K-band magnitude of the nova source. Simultaneously, we obtained high resolution spectra covering $4000-9800 \AA$ using the Hamilton Spectrograph and Coud\'e Auxilliary Telescope (CAT) at Lick Observatory.\\
 Finally, combining interferometric observations with radial expansion velocities obtained from line profiles allows us to derive the distance to the nova. The results can then be compared with expansion distances found from photometric analysis.

\section{Observations and data reduction}

\subsection{Interferometric observations at PTI }
\begin{figure*}[!ht]
\plotone{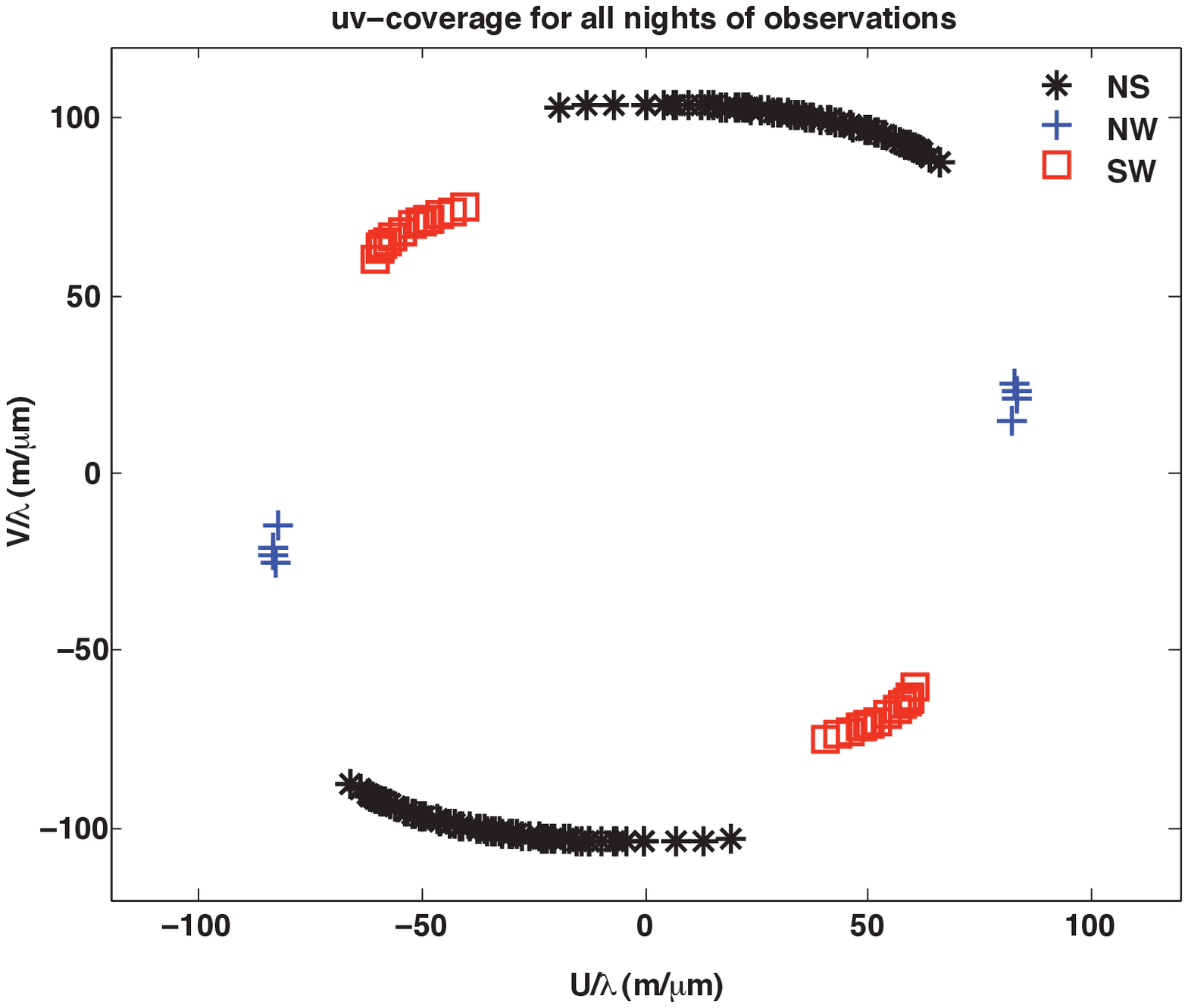}
\caption[UV COVERAGE]
{ \label{fig::uvplane}
\textit{uv}- plane coverage of observations of nova V458 Vul in spatial frequencies.
}
\end{figure*}

Nova V458 Vul was observed in the NIR at 2.2 $\rm {\mu m}$ wavelength for nine nights between 2007 August 10 and 2007 August 21. On each of these nights observations were conducted using the North-South (NS) baseline. On four of these nights, data were obtained at a second baseline: one night with the North-West (NW) and the rest using the South-West (SW) baseline. Figure \ref{fig::uvplane} shows \textit{uv}-plane coverage of the observations and a summary of the observations can be found in Table \ref {tab::obslog} (the complete table is available in the electronic version).\\

The squared visibility measured by PTI were calibrated by dividing the target visibility by the system visibility $\rm {V^{2}}_{sys}$, the response of the interferometer to an unresolved point source. The result is a calibrated target visibility which would have been measured with an ideal interferometer without any loss of coherence. The data reduction was performed using webCalib (http://nexsci.caltech.edu/software/V2calib/wbCalib/index.html) and calibrators have been chosen with getCal (http://nexsciweb.ipac.caltech.edu/gcWeb/gcWeb.jsp). WebCal and getCal are provided by the NASA Exoplanet Science Institute. For further details, see \cite{Colavitaetal1999}. Properties of calibrators are given in Table \ref{tab::calib}. 

\begin{deluxetable}{lllllll}
\tablecolumns{7}
\tablewidth{0pc} 
\tablecaption{Log of the interferometric observations.  \label{tab::obslog}}
\tablehead{ 
\colhead{Date (MJD)} &\colhead{ $\lambda (\mu m)$}& \colhead{BL} &  \colhead{U(m)} &  \colhead{V(m)} &\colhead{V$^{2}$} & \colhead{$\sigma_{V^{2}}$}}
\startdata
54322.16426    &2.2402	&NS  &    -65.976951    	&  -87.392875       &     0.934857&	0.105594	\\
54322.17632	& 2.251     &NS  &	-63.974559	& -89.153965      & 0.982534&	0.0980614\\	
54322.20719	& 2.2308     &NS  &	-57.197144	& -93.371204      & 0.634843&	0.114107	\\
54322.22091    &2.2286	&NS  &	-53.472840	& -95.079388      & 0.920575	&      0.15715	\\
54322.25196    &2.2326	&NS  &	-43.621484	& -98.478271      & 0.770948	&     0.0936088	\\
\enddata
\tablecomments{Table \ref{tab::obslog} is published in its entirety in the electronic edition of \emph{Astrophysical Journal}. A portion is shown here for guidance regarding its form and content.}
\end{deluxetable}

\begin{deluxetable}{llllll}
\tablecolumns{6}
\tablewidth{0pc} 
\tablecaption{Parameters of the calibrators  \label{tab::calib}}
\tablehead{ 
\colhead{Calibrator} & \colhead{V} & \colhead{K} & \colhead{Spectral Type} & \colhead{Diameter (mas)} & \colhead{Separation(deg)}}
\startdata
HD 180242 &    6.04 &      4.05            &     G8 III   & 0.7        & 9.2       \\
HD 190406 &    5.80       &              4.39    &   G0 V      & 0.6        & 4.4     
\enddata
\tablecomments{ In the last column, separation denotes the angular distance from the calibrator to the nova.}
\end{deluxetable}

\subsection{Spectroscopy}
During the early days of explosion in August 2007, we obtained high dispersion wide bandpass spectra of the nova from the Hamilton Echelle Spectrometer using the Coud\'e Auxilliary Telescope at Lick Observatory. The Hamilton Echelle Spectrometer, installed at the coude focus of the Shane 3-m telescope, is a high dispersion spectrograph  designed primarily for high resolution (R =20000 - 60000) wide bandpass spectroscopy of point-like sources down to a limiting magnitude of about V = 16.5, over the 0.34 $\mu$m to 1.1 $\mu$m spectral region \citep{Vogt1987}.  The Hamilton can also be fed by the 0.6 m Coude Auxilliary Telescope (CAT), which was used for all the spectra of the nova in our study. This instrument was available to one of the co-authors working on TATOOINE project \citep{Konacki2009}, when the nova was announced. The journal of spectroscopic observations is presented in Table \ref{tab::specobs}. The dispersion ranges from 0.03 to 0.09 $\AA$/pixel at 4000 to 9800 $\AA$, respectively (about 0.053 at 6500 $\AA$). The Hamilton reduction procedure involves using several image cleaning techniques before extraction of the spectra. To clean the image,
the areas of the image that are affected by cosmic rays must be identified.
First, a local median image is created by passing the original image
through a median filter with a 5x5 pixel kernel.
All pixels of the original image with an intensity value greater than
2.85 $\sigma$ of the median value are flagged as cosmic ray hits and
then replaced by the median value of the neighboring pixels.\\

To remove small scale pixel--to--pixel variations in CCD sensitivity,
a localized flat field correction is applied.
Using a sum of 50 continuum-source flat field images,
the location and width of each order is determined.
Polynomials are fit to the curved orders and then the orders
are straightened by extracting an odd number of pixels centered around the
polynomial fits.\\
A normalized flat field image is then created for each order
by dividing each pixel of the raw flat field image by a local median value
(determined by applying a median filter with a kernel of 25 pixels
along the dispersion direction only).
The resulting flat field images all have an average value nearly equal to one.
Each order of the science image is divided by the corresponding flat field
image.\\
A simple extraction procedure is employed to create the source and
detector background spectra for each order. A slit profile is
created by collapsing each order along the dispersion direction.
The science spectrum is
the sum of all rows greater than 7\% of the maximum slit profile value, while
the detector background is chosen to be the mean of all rows less than 3\%
of the maximum slit profile value. To remove Poisson variations from the
mean background spectra, they are median filtered with a kernel of 15 pixels.
The mean background spectra are then subtracted from the source spectra
to create the final science spectra.

\begin{deluxetable}{llll}
\tablecolumns{4}
\tablewidth{0pc} 
\tablecaption{Hamilton HIRES Spectroscope Observation Log \label{tab::specobs}}
\tablehead{ 
\colhead{Date (MJD)}&\colhead{Date (UT)} & &\colhead{SNR (at H$\alpha$)}}
\startdata
54322  &2007 Aug 10 6:21  & & 17.5\\
54323  &2007 Aug 11 5:14   & &  22.4   \\
54325  &2007 Aug 13 5:29     & & 38.0  \\
54327  &2007 Aug 15 5:18      &  & 21.8
\enddata
\tablecomments{Following the observing dates, starting times for each exposure of 2100 sec are given.}
\end{deluxetable}

\subsection{Photometry}

\begin{figure*}[!ht]
\plotone{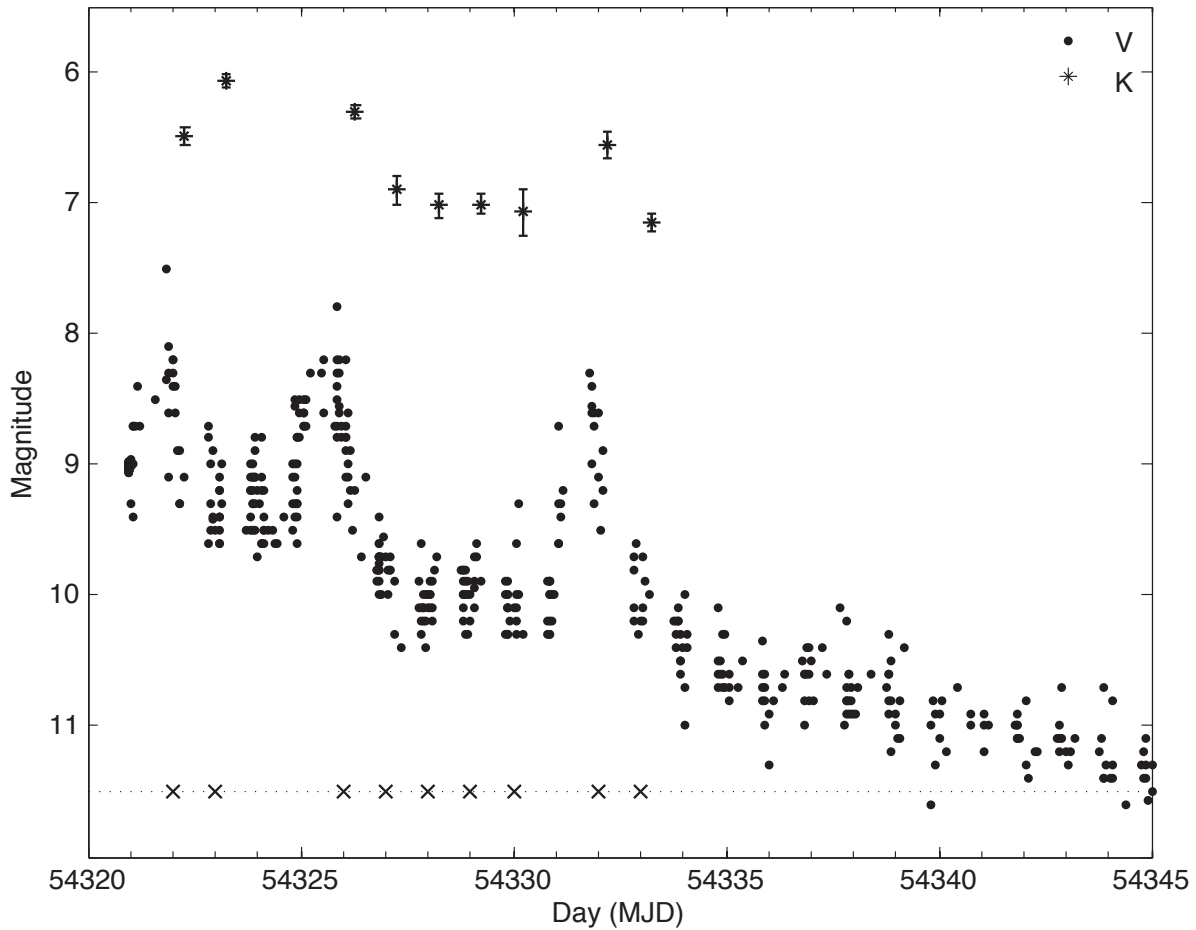}
\caption[PHOTOMETRY CURVE]
{ \label{fig::photometry}
Infrared and Visible light curve of the nova taken by PTI (K band) and AAVSO international database (V band). Crosses at the bottom of the figure denote the days at which interferometric observations were conducted.
}
\end{figure*}

As described in \citep{Colavitaetal1999} PTI is equipped with both a broadband fringe tracking channel and a 5-channel low-resolution spectrometer covering the K band.  In addition to the fringe visibilities, photon count rates are recorded and can be used for low-precision photometry.\\
We used the recorded photon count rates in the broad-band channel together with 2MASS K-band magnitude estimates of the calibrator sources \citep{Beichmann88} to estimate the apparent K-band magnitude of the nova source on a nightly basis.  It should be remembered that PTI was not designed for high-precision photometry and hence the K-magnitudes, while useful, should be treated with caution.\\
We have also used the AAVSO photometric database to collect photometric data for our target (http://www.aavso.org). Figure \ref{fig::photometry} shows both infrared and visible light magnitude evolution of the nova from the early days of the explosion. This unusual nova shows several episodes of rebrightenings in both photometric bands . 

\section{Results and modeling}
\begin{figure*}[!ht]
\plotone{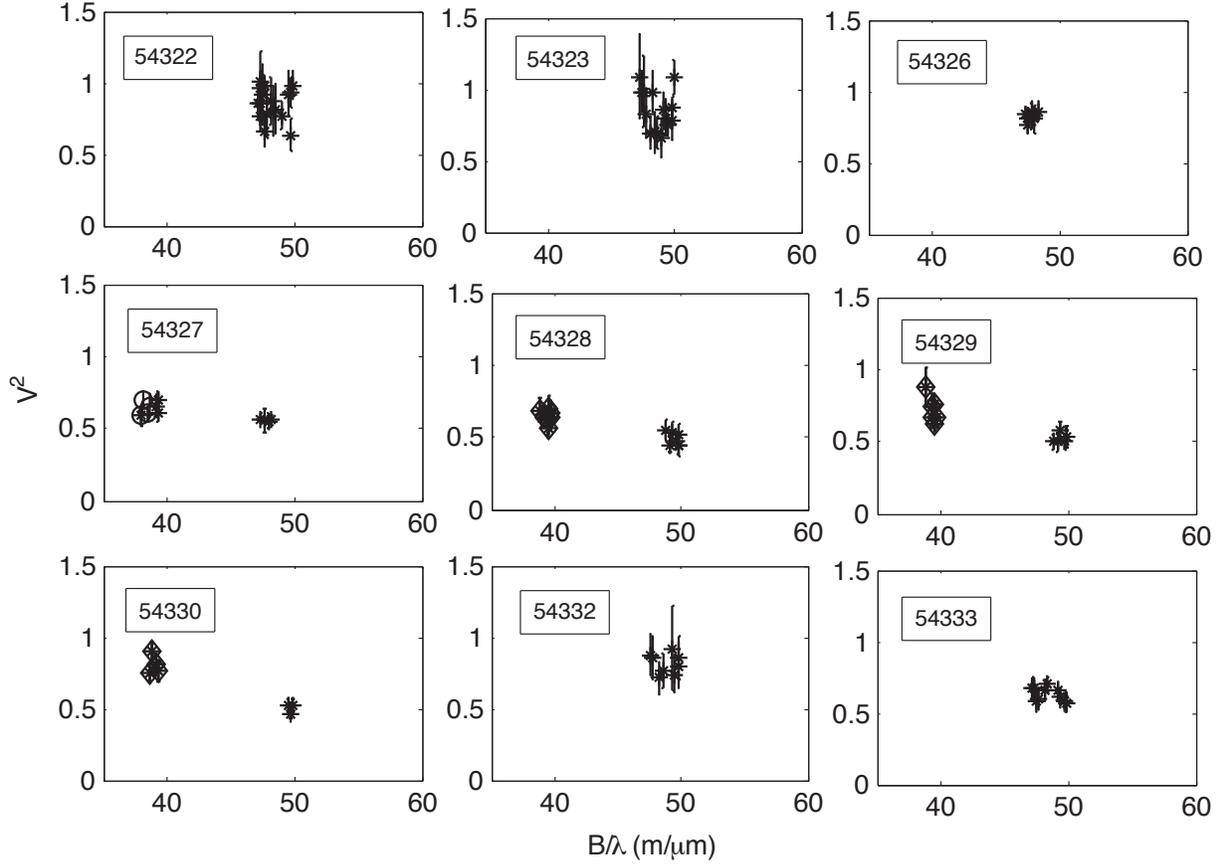}
\caption[OBSERVED VISIBILITY]
{ \label{fig::observedvis}
Squared visibilities versus spatial frequencies. Stars, circles and diamonds symbols correspond to observations at NS, NW and SW baselines. Observation dates (MJD) are shown on a box on each graph.
}
\end{figure*}

In this section we present a summary of the results followed by analysis of the results using several emission models and present a model fit based on parameters including size, inclination, position angle and aspect ratio. Figure \ref {fig::observedvis} shows visibilities as a function of spatial frequency B/$\lambda$.

Visibility measurements with values below unity imply that the nova is spatially resolved. The van Cittert-Zernike theorem provides a mathematical description of the relation between source brightness distribution and fringe visibility. For a circularly symmetric uniform disk characterized only by its diameter $\theta_{UD}$, the squared visibility is given by \citep{Lawson1999}

\begin{equation}
V^{2}=  {\Bigl[\frac{2J_{1}(\pi \theta_{UD}r_{\textit{uv}})}{\pi \theta_{UD}r_{\textit{uv}}}  \Bigr]}^{2}
\end{equation}

where $J_{1}$ is the first-order Bessel function and $r_{uv}=\sqrt{\textit{u}^2+\textit{v}^2}$; with \textit{u} and \textit{v} as projected baselines where
\begin{equation}
\frac{\vec{B}.\vec{s}}{\lambda}=\sqrt{u^2+v^2}
\end{equation}
$\vec{B}$ is the baseline of the observation, $\vec{s}$ is a unit vector pointing toward the source and $\lambda$ is the observing wavelength. Allowing the disk to be inclined (or, equivalently, elliptical) adds two more free parameters to a simple uniform disk model; inclination $\phi$ and position angle $\psi$ which will change spatial frequencies accordingly as follows
\begin{eqnarray}
u^{\prime}&=&u \sin\psi+v \cos\psi \\
v^{\prime}&=&\cos\phi(v \sin\psi-u \cos\psi)
\end{eqnarray}

For a face-on Gaussian distribution of the emission with diameter $ {\theta_{Gauss}}$, visibility is given by
\begin{equation}
V= e^{\Big(-\frac{\pi^{2}{\theta^{2} _{Gauss}}{r^2_{uv}}}{4\ln 2}\Big)}
\end{equation}
Inclination can also be added to the face-on Gaussian model using the inclined spatial frequencies discussed above.

In the first part of our model fitting we assumed that more of the NIR emission originates in the circumstellar region of the erupted nova than from the central star. This simplifies visibility modeling as it reduces the emitting components to one. This is justified because the system's luminosity increased by $>5$ mag during outburst, implying the central star is at least $100\times$ fainter than the outburst material.\\
We performed least-squared fits of a face-on uniform disk, inclined disk, face-on Gaussian and inclined Gaussian models to the observed visibilities. For each model, all parameters for all nights are determined simultaneously using a non-linear fit. A different size is allowed for each night, but the position angle and inclination (if included in the model) are assumed to be identical from one night to the next. The best-fit parameters of different models along with reduced $\chi^{2}_{r}$ values are listed in Table \ref{tab::allmodel}. An extended emission disk can successfully represent our interferometric observations.

Next, we tried to fit the observed visibilities with a model consisting of two components, a resolved source (a disk) and an unresolved one (a point source resembling the central star). For such a model, the visibility is given by:

 \begin{eqnarray}
V=\Bigl( \frac{1+ rV_{D}}{1+r} \Bigl)
\end{eqnarray}
where $V_{D}$ is the visibility of the disk component and $r$ is the intensity ratio of the resolved and the unresolved components.

First we took the uniform disk as the resolved emitting component. We kept the disk size fixed at each night and let the intensity ratio $r$ change from night to night. We tried this model for both face-on and inclined geometries. Both models do not produce a satisfactory fit. We also used both face-on and inclined Gaussian disk as the extended emitting component and we still could not find a satisfactory fit to observations.
The next effort was to use a two-component model, but varying both the size of extended emission and  intensity ratio at each night. We used both uniform and Gaussian disk and for both face-on and inclined geometries.  All these models resulted in too many degrees of freedom and therefore none of the parameters could be constrained and thus the fit failed.

\begin{deluxetable}{ll}
\tablecolumns{2}
\tablewidth{0pc} 
\tablecaption{V458 Vul model fitting parameters  \label{tab::allmodel}}
\tablehead{ 
\colhead{Date (MJD)} & \colhead{$\theta$ (mas)}}
\startdata
\multicolumn{2}{c}{Face- on UD model, $\chi^{2}_{r}=0.834$} \\
\hline
 54322   & 1.063 $\pm$ 0.117       \\ 
  54323        &1.217  $\pm$ 0.114    \\ 
 54326         & 1.223$\pm$0.098  \\  
   54327      &2.087$\pm$0.054    \\  
  54328      &2.246     $\pm$0.060    \\ 
  54329       &2.083     $\pm$ 0.051    \\ 
  54330   & 2.040      $\pm$ 0.058  \\ 
  54332      & 1.301    $\pm$ 0.175  \\ 
    54333     &1.820     $\pm$0.065     \\  
\hline
 \multicolumn{2}{c}{Face-on Gaussian, $\chi^{2}_{r}=0.845$} \\
\hline
  54322   &0.630$\pm$0.070    \\ 
 54323     & 0.722$\pm$0.069    \\  
   54326     &  0.726$\pm$0.059  \\ 
   54327   & 1.259$\pm$0.034    \\  
   54328    & 1.360$\pm$0.039   \\  
  54329   & 1.256$\pm$0.033    \\ 
 54330  & 1.228$\pm$0.037    \\  
  54332   & 0.773$\pm$0.116    \\ 
    54333    & 1.093$\pm$ 0.041  \\ 
 \hline
 \\
   \multicolumn{2}{c}{Inclined UD model, $\chi^{2}_{r}=0.773$} \\
     \multicolumn{2}{c}{$\phi$= 29.0$\pm$ 6.0 ,$\psi$= 70.6 $\pm$ 10.2} \\
\hline
  54322 & 1.161        $\pm$ 0.139     \\ 
 54323 &  1.294       $\pm$0.132   \\ 
  54326 &   1.345        $\pm$ 0.131  \\ 
   54327&  2.242        $\pm$ 0.106  \\ 
  54328&  2.412       $\pm$  0.111    \\ 
  54329&    2.232     $\pm$0.097   \\ 
 54330&   2.190     $\pm$0.104    \\ 
   54332 &     1.393     $\pm$0.198  \\ 
    54333&    1.964      $\pm$0.116    \\
   \hline
   \\
  \multicolumn{2}{c}{Inclined Gaussian,  $\chi^{2}_{r}=0.772$} \\
 \multicolumn{2}{c}{$\phi$= 29.7 $\pm$ 5.8, $\psi$=67.5 $\pm$ 10.9}\\   
\hline
  54322 &  0.687        $\pm$ 0.083    \\ 
  54323 &  0.765       $\pm$0.079  \\ 
   54326 &   0.780       $\pm$ 0.078  \\ 
   54327&  1.349        $\pm$ 0.064  \\ 
  54328&  1.463      $\pm$  0.068   \\ 
  54329&    1.348     $\pm$ 0.059\\ 
  54330&   1.319     $\pm$  0.064  \\ 
   54332 &     0.825     $\pm$0.119\\ 
    54333&     1.175      $\pm$ 0.070  \\
   \enddata
\tablecomments{$\theta$ refers to the size of angular diameter for face-on models and major axis diameter for inclined models, $\phi$ is the inclination and $\psi$ is the position angle. }
\end{deluxetable}

\begin{figure*}[!ht]
\plotone{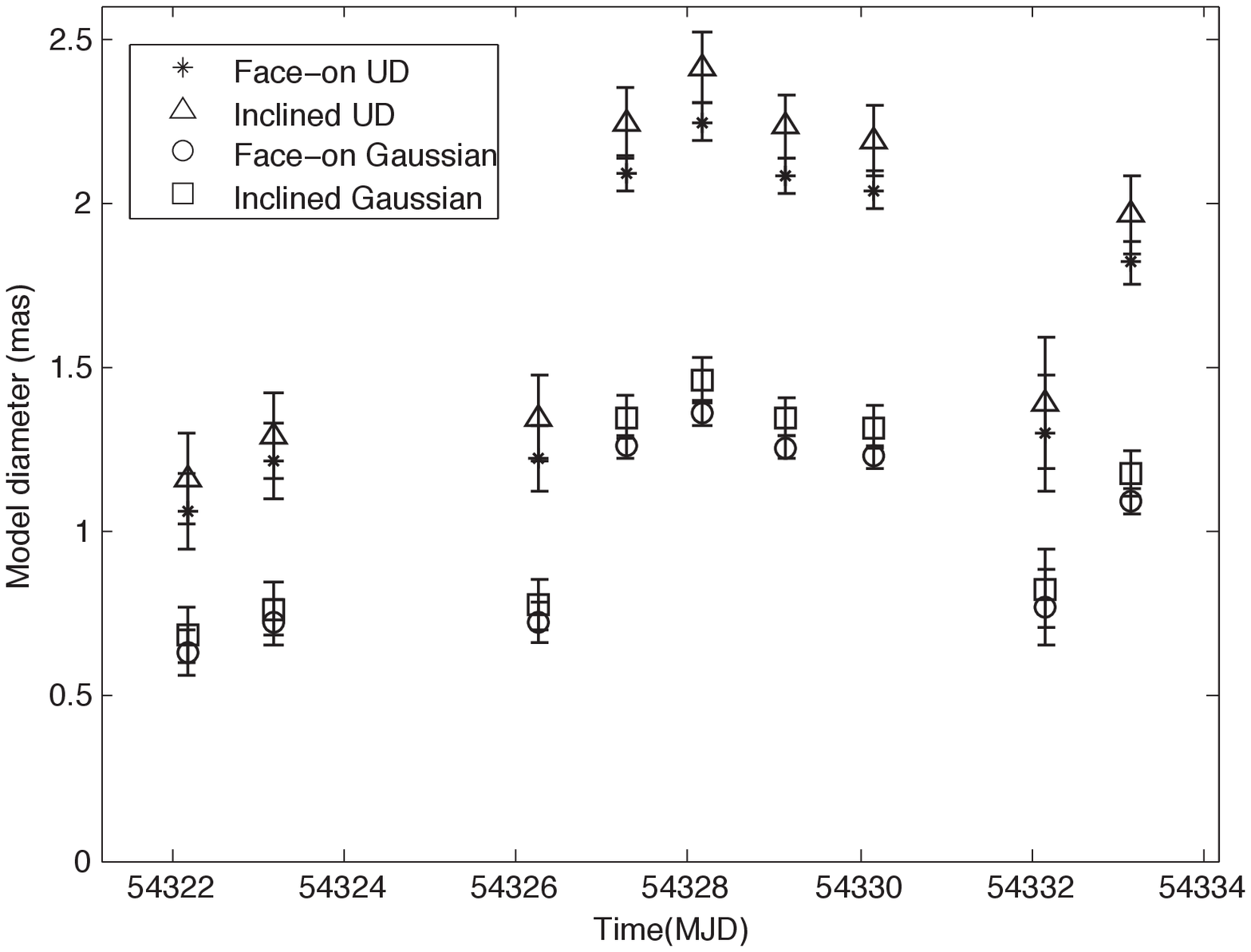}
\caption[MODEL :SIZES]
{ \label{fig::sizes}
The best-fit of angular sizes of all models as a function of observation date.}
\end{figure*}

Among the single component models, the inclined uniform disk and Gaussian models give lower Chi-squared values compared to face-on models which makes the inclined models preferable for the V458 Vul system. Both inclined models give similar inclination and position angle of $\approx 30$ and $\approx 70$ degrees respectively. The inclination found for the system is not very high though, which explains the small difference in angular sizes found for the nova using face-on and inclined models. The inclination can be interpreted as $\cos\phi$ being an aspect ratio of an elliptical disk. The aspect ratios are $0.875\pm0.055$ and $0.869\pm0.055$ for the uniform disk and Gaussian models, respectively.
As it can be seen in Figure \ref{fig::sizes}, for all models, the disk diameter does not follow a constant trend, but it shows some small fluctuations of decrease and increase in our observation period at the early days after the explosion. The size fluctuation is consistent with the rebrightening photometric variations seen in the V458 Vul light curve in Figure \ref{fig::photometry}. When the flux increases but the temperature remains constant, the size of the emitting region decreases and vice versa. The only exception is at the very beginning of the outburst from day 54322 to day 54323 during which the size of the emission increases and at the same an increase in K flux is observed. This behavior can not be explained by current observations.

\begin{figure*}[!ht]
\plotone{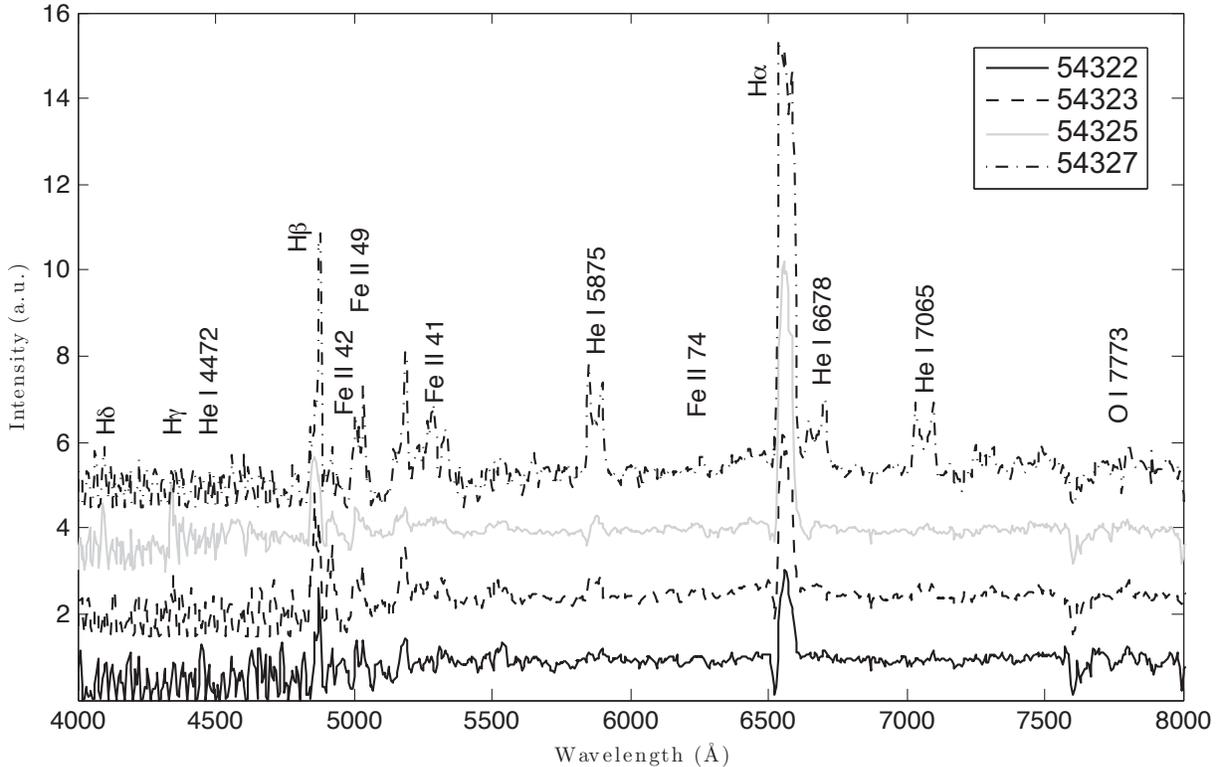}
\caption[MODEL :SPECTRA]
{ \label{fig::novaspectra}
V458 Vul2007 spectra. An offset is given to the plots of different  observing days. }
\end{figure*}

\begin{figure*}[!ht]
\plotone{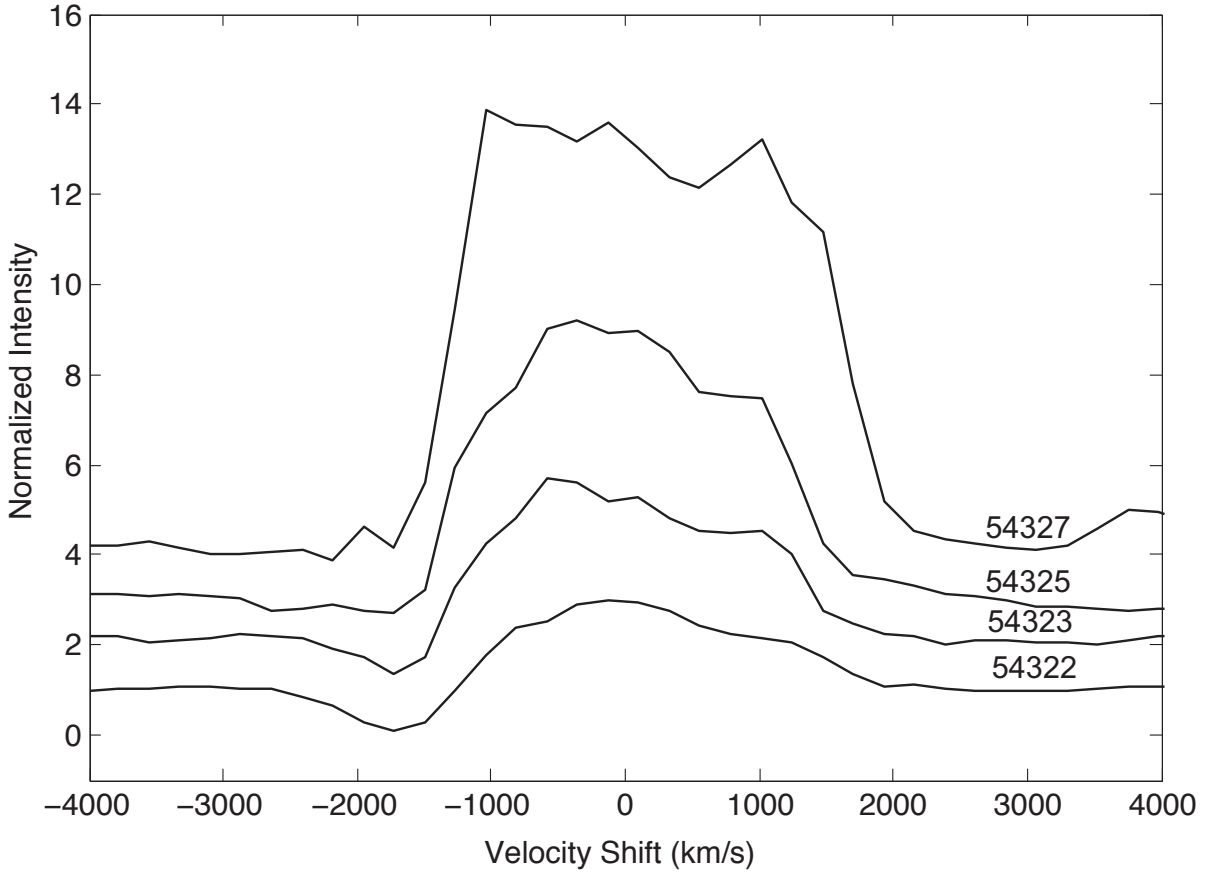}
\caption[MODEL :HALPHA]
{ \label{fig::balmerlines}
H$\alpha$ line profile during our observation period. An offset of $1.5$ units is given to the plots of different  observing days. }
\end{figure*}

The observed spectra of the nova, illustrated in Figure \ref{fig::novaspectra}, shows strong Balmer lines at the early days of the explosion. Most of the emission lines, including Balmer lines and the He I line change significantly at different epochs of observation, indicating the fast varying nature of the ejected material. The evolution of the H$\alpha$ line is shown in Figure \ref{fig::balmerlines}. The emission line undergoes rapid changes from day 54322, one day after outburst, to day 54327, when 6 days had elapsed since the nova explosion. The H$\alpha$ line shows a P-Cygni absorption feature indicative of optically thick outflowing gas at a velocity of $\approx$ 1700 km/s. However, the absorption decreases as the nova evolves, which can be interpreted as a transition of the ejecta to a more optically thin regime. \cite{Tanaka2011} present evolution of V458 Vul spectra during ten nights following the outburst. They also observe a strong P-Cygni feature for the H$\alpha$ line immediately after outburst and relate the disappearance of this absorption feature in Balmer lines to rebrightenings in the nova and they predict the reapperance of those features in the next nova brightness maximums after the initial one. Unfortunately we could not confirm this since our spectroscopic observations are only until day 54327, six days after outburst while the first rebrightening occurs eleven days after the explosion. We also measured FWHM of H$\alpha$ lines at different epochs. The H$\alpha$ FWHM at days 54322 and 54323 when the absorption feature is present, are 2000$\pm 50$ km/s and 2350$\pm 50$ km/s respectively, while the P-Cygni absorption presents a wind velocity of $\approx$ 1700 km/s.\\

\section{Discussion}
We modeled 2.2 $\mu$m emission from the classic nova V458 Vul 2007 using several models including single component and two-component emitting models for uniform and Gaussian disks, and for both face-on and inclined geometries. A single extended emission model can check if the origin of most of the emission is in the ejected wind, while using a two-component model probes whether the photosphere of the White Dwarf can significantly change the visibilities and thus the inferred emission sizes. A single extended emission component can successfully represent the current observations while none of two-component models can produce a satisfactory fit. However by considering an extended emission disk it is difficult to explain the origin of decreasing size of the emission after day 54328 and increasing size at day 54333 (See Figure \ref{fig::sizes}) due to rebrightenings.\\
The low inclination of the system, or equivalently, the elliptical aspect ratios being near unity, found for V458 Vul is compatible with the nova being in the early fireball phase of the explosion in which the hot ejected gas is seen as an expanding photosphere with spherical geometry cooling adiabatically \citep{Gehrzetal1998}. However, we emphasize that more multibaseline observation of the nova would have been needed to investigate the symmetry of this system. It does, however, give initial insights into how some novae evolve.\\


The nova light curve, shown in Figure \ref{fig::photometry}, does not follow the simple visual light curve expected for classical novae, in which the light rises to a maximum and then declines until the nova goes into the post-outburst state, where the light remains constant \citep{McLaughlin1960}. Instead it shows a few peaks before the decline phase starts up. These oscillations are not very common in the initial stage of novae explosions and may be caused by shocks originating from the interaction of the ejecta with the surrounding nebula, which has been observed around the nova central source. These shocks are the most plausible sources of the hard X-ray emission detected in this nova \citep{Ness2009}.\\

\begin{figure*}[!ht]
\plotone{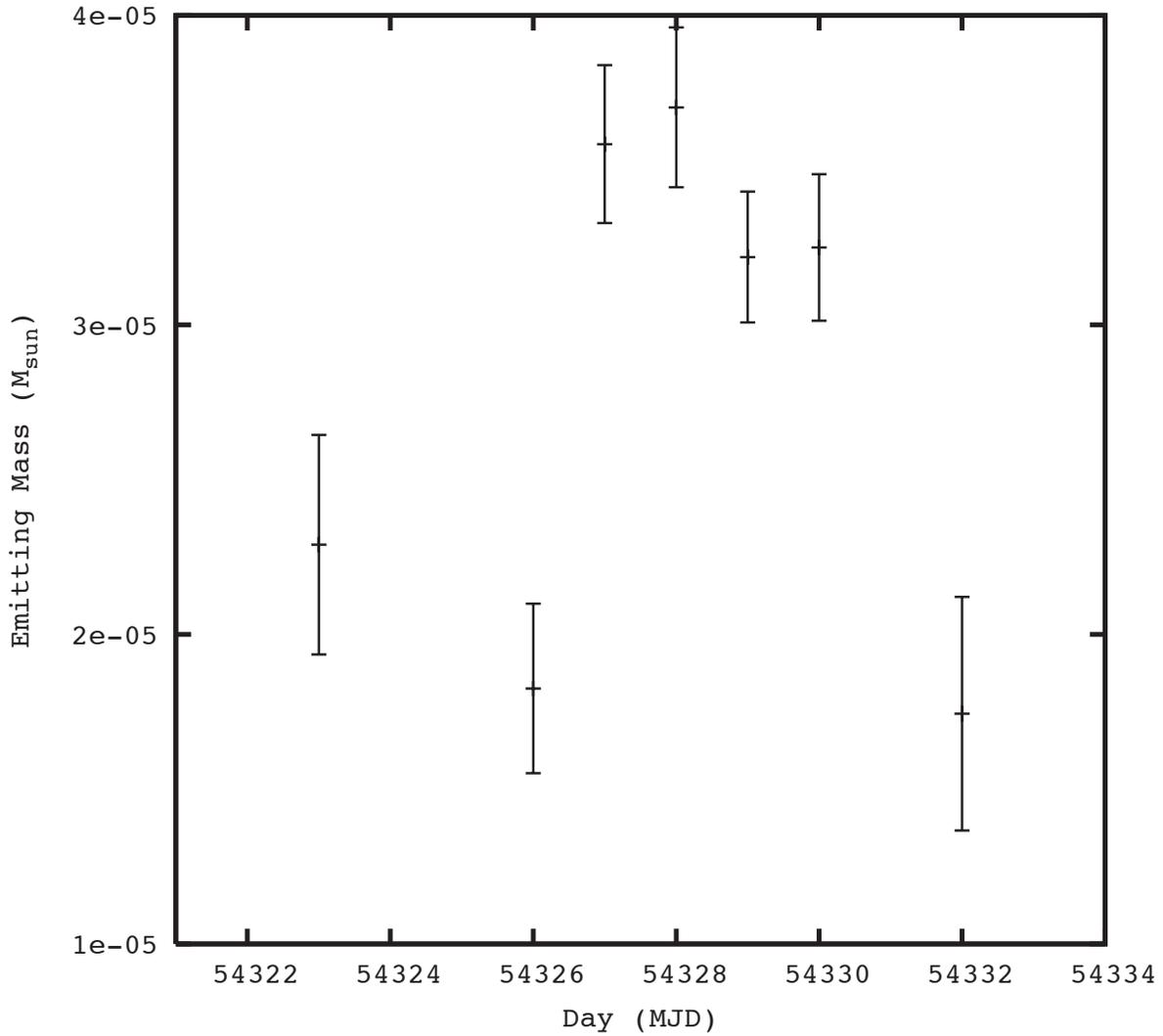}
\caption[OBS :MASS]
{ \label{fig::novaemitmass}
The emitting mass as a function of time measured by PTI. The best fit Gaussian angular diameters are used to compute the emitting volume. }
\end{figure*}

We also used our K photometry and reddening estimate of E(B-V)=0.63 \citep{Wessonetal2008} to compute the nova K-band flux. 
Using the measured flux and assuming a temperature of $\approx 10^{4} K$ \citep{Starrfield1987} we computed the emitted intensity and thus the electron density and emitting mass as a function of time. As prescribed by \cite{Lane07} we assumed the emitting region to be an ellipsoid  with dimensions of the best-fit Gaussian model and with a uniform electron density distribution. The emitting mass as a function of time is shown in Figure \ref{fig::novaemitmass}. The ejected mass is between $1-4\times 10^{-5} M_{\odot}$. Given the duration of mass ejection ($\approx$ 415 days; \cite{Hachisu2010} ) and the expansion velocity of  $\approx$ 1700 km/s,  we estimated the total ejected mass of $\approx 4\times 10^{-4} M_{\odot}$, which is consistent with the amount expected for Classical novae ($\approx 10^{-4} M_{\odot}$; \cite{Prialniketal1995}).\\

One of the important questions regarding novae systems is at what distance they occur. An accurate distance to a nova allows modeling of the accretion process in the post-nova system without a distance constraint \citep{Wade1998}. Several methods have been used to estimate the distance to the Nova V458 Vul. Using absolute magnitude relation for computing the distance, exploiting MMRD calibration versus decline times $t_2$ and $t_3$, gives the distance of 6.7-10.3 kpc to this Nova \citep{Poggiani2008}.  In a different study, the nova distance is estimated as 13 kpc, based on the light traveling time for the nova to reach the inner nebular knot \citep{Wessonetal2008}. The latter distance is supported by another estimation of the nova distance using the nebular radial velocity measurement by the same authors. Interferometric observations can also be used to estimate the distance to novae \citep{Lane05,Lane07}. During the early days of the explosion during which the diameter of the emission is increasing, the angular expansion rate can be used to find the onset of the explosion. If we assume that the nova shell is expanding spherically symmetric,  using the expansion velocity from spectroscopy, the nova size can be computed. The nova distance can then be computed from the nova shell linear and angular sizes. This is very straightforward for the uniform disk models and it gives a distance and formal uncertainty of $9.9\pm0.3$ kpc for V458 Vul. However for the case of the inclined disk, the emission is coming from a prolate spheroid and the inclination effect should be taken into account. Following the prescription by \cite{Wade2000} we computed the nova distance for the elliptical disk and found a distance and uncertainty of $11.4\pm0.5$ kpc. Both distances derived from these models are consistent with the range (6.7-13 kpc) derived so far for this nova but have much better precision.
The observed spectra show Fe II lines and Balmer lines immediately after the outburst which places the nova in the FeII class of novae \citep{Williams1992} as has also been suggested by \cite{Poggiani2008}. Only a few days after the outburst, He II lines appear, placing the nova in He/N class of novae with white dwarfs of higher masses having high velocity shells of ejecta. This change of novae class happens even faster than what has been expected before and is indicative of the very fast varying and complicated nature of this nova. Different shapes and velocities of emission lines in the nova show that they arise in different parts of the gas being ejected and at different speeds and directions. The detailed analysis of the line profiles will be realized in a subsequent paper.

\section{Conclusions}

We used PTI to resolve 2.2 $\mu$m emission from the very fast classical nova V458 Vul 2007. Our analysis shows that the nova ejecta has a geometry with low inclination (aspect ratio near unity) which is supported with the nova being in the fireball phase at which the outflowing gas is optically thick, confirmed by the presence of strong P-Cygni Balmer lines in the spectra. The expansion velocity derived from the P-Cygni profile of the H$\alpha$ line is 1700 km/s. This expansion velocity, along with the rate of angular size change, gives a distance of 9.9-11.4 kpc to the nova, which is consistent with the distances estimated for this nova using other techniques. The very quick transition of the nova from Fe II to He/N class  within a few days after its outburst reveals the fast varying complicated nature of V458 Vul. The total ejected mass from the nova is $\approx 4\times 10^{-4} M_{\odot}$. Future work will be focused on studying abundances of different species observed in the spectra of V458 Vul in the first few days after explosion. This will help uncover the mystery of the irregular behavior of V458 Vul and will provide some constraints on the evolution of fast binaries.

\section{Acknowledgments}

MWM acknowledges  support from a UC Berkeley Towner postdoctoral fellowship and an internal UC Berkeley  SSL grant that funded the acquisition of spectroscopic data. SR and MWM acknowledge the support from Center of Excellence of Tennessee State University. AG acknowledges support from NSF grant 0958267. We thank Don Neill for fruitful discussions. We are also
grateful to our referee, Olivier Chesneau, for useful comments that
improved the manuscript.
\bibliography{main}

\begin{thebibliography}{44}
\providecommand{\natexlab}[1]{#1}
\providecommand{\url}[1]{\texttt{#1}}
\expandafter\ifx\csname urlstyle\endcsname\relax
  \providecommand{\doi}[1]{doi: #1}\else
  \providecommand{\doi}{doi: \begingroup \urlstyle{rm}\Url}\fi

\bibitem[{Arai} et~al.(2009){Arai}, {Uemura}, {Kawabata}, {Sasada}, {Ohsugi},
  {Yamashita}, {Sato}, {Kino}, and {Kanata Team}]{Arai2009}
A.~{Arai}, M.~{Uemura}, K.~S. {Kawabata}, M.~{Sasada}, T.~{Ohsugi},
  T.~{Yamashita}, S.~{Sato}, M.~{Kino}, and {Kanata Team}.
\newblock \emph{The Eighth Pacific Rim Conference on Stellar Astrophysics: A
  Tribute to Kam-Ching Leung}, volume 404.
\newblock Astronomical Society of the Pacific Conference Series, 2009.

\bibitem[{Barry} et~al.(2008){Barry}, {Danchi}, {Traub}, {Sokoloski},
  {Wisniewski}, {Serabyn}, {Kuchner}, {Akeson}, {Appleby}, {Bell}, {Booth},
  {Brandenburg}, {Colavita}, {Crawford}, {Creech-Eakman}, {Dahl}, {Felizardo},
  {Garcia}, {Gathright}, {Greenhouse}, {Herstein}, {Hovland}, {Hrynevych},
  {Koresko}, {Ligon}, {Mennesson}, {Millan-Gabet}, {Morrison}, {Palmer},
  {Panteleeva}, {Ragland}, {Shao}, {Smythe}, {Summers}, {Swain}, {Tsubota},
  {Tyau}, {Wetherell}, {Wizinowich}, {Woillez}, and {Vasisht}]{Barry2008}
R.~K. {Barry}, W.~C. {Danchi}, W.~A. {Traub}, J.~L. {Sokoloski}, J.~P.
  {Wisniewski}, E.~{Serabyn}, M.~J. {Kuchner}, R.~{Akeson}, E.~{Appleby},
  J.~{Bell}, A.~{Booth}, H.~{Brandenburg}, M.~{Colavita}, S.~{Crawford},
  M.~{Creech-Eakman}, W.~{Dahl}, C.~{Felizardo}, J.~{Garcia}, J.~{Gathright},
  M.~A. {Greenhouse}, J.~{Herstein}, E.~{Hovland}, M.~{Hrynevych},
  C.~{Koresko}, R.~{Ligon}, B.~{Mennesson}, R.~{Millan-Gabet}, D.~{Morrison},
  D.~{Palmer}, T.~{Panteleeva}, S.~{Ragland}, M.~{Shao}, R.~{Smythe},
  K.~{Summers}, M.~{Swain}, K.~{Tsubota}, C.~{Tyau}, E.~{Wetherell},
  P.~{Wizinowich}, J.~{Woillez}, and G.~{Vasisht}.
\newblock \emph{\apj}, 677:\penalty0 1253--1267, 2008.

\bibitem[{Beichmann}(1988)]{Beichmann88}
C.~A. {Beichmann}.
\newblock \emph{The Study of Star Formation with IRAS}, volume 297 of
  \emph{Lecture Notes in Physics, Berlin Springer Verlag}.
\newblock Berlin Springer Verlag, 1988.

\bibitem[{Bianciardi} et~al.(2007){Bianciardi}, {Villegas}, and
  {Sanchez}]{Bianciardi2007}
G.~{Bianciardi}, J.~M. {Villegas}, and A.~{Sanchez}.
\newblock \emph{Central Bureau Electronic Telegrams}, 1035:\penalty0 2, 2007.

\bibitem[{Bode}(2010)]{Bode2010}
M.~F. {Bode}.
\newblock \emph{Astronomische Nachrichten}, 331:\penalty0 160, 2010.

\bibitem[{Chesneau} et~al.(2007){Chesneau}, {Nardetto}, {Millour}, {Hummel},
  {Domiciano de Souza}, {Bonneau}, {Vannier}, {Rantakyr{\"o}}, {Spang},
  {Malbet}, {Mourard}, {Bode}, {O'Brien}, {Skinner}, {Petrov}, {Stee},
  {Tatulli}, and {Vakili}]{Chesneau2007}
O.~{Chesneau}, N.~{Nardetto}, F.~{Millour}, C.~{Hummel}, A.~{Domiciano de
  Souza}, D.~{Bonneau}, M.~{Vannier}, F.~{Rantakyr{\"o}}, A.~{Spang},
  F.~{Malbet}, D.~{Mourard}, M.~F. {Bode}, T.~J. {O'Brien}, G.~{Skinner}, R.~G.
  {Petrov}, P.~{Stee}, E.~{Tatulli}, and F.~{Vakili}.
\newblock \emph{\aap}, 464:\penalty0 119--126, 2007.

\bibitem[{Chesneau} et~al.(2008){Chesneau}, {Banerjee}, {Millour}, {Nardetto},
  {Sacuto}, {Spang}, {Wittkowski}, {Ashok}, {Das}, {Hummel}, {Kraus},
  {Lagadec}, {Morel}, {Petr-Gotzens}, {Rantakyro}, and
  {Sch{\"o}ller}]{Chesneau2008}
O.~{Chesneau}, D.~P.~K. {Banerjee}, F.~{Millour}, N.~{Nardetto}, S.~{Sacuto},
  A.~{Spang}, M.~{Wittkowski}, N.~M. {Ashok}, R.~K. {Das}, C.~{Hummel},
  S.~{Kraus}, E.~{Lagadec}, S.~{Morel}, M.~{Petr-Gotzens}, F.~{Rantakyro}, and
  M.~{Sch{\"o}ller}.
\newblock \emph{\aap}, 487:\penalty0 223--235, 2008.

\bibitem[{Chesneau} et~al.(2011){Chesneau}, {Meilland}, {Banerjee}, {Le
  Bouquin}, {McAlister}, {Millour}, {Ridgway}, {Spang}, {Ten Brummelaar},
  {Wittkowski}, {Ashok}, {Benisty}, {Berger}, {Boyajian}, {Farrington},
  {Goldfinger}, {Merand}, {Nardetto}, {Petrov}, {Rivinius}, {Schaefer},
  {Touhami}, and {Zins}]{Chesneau2011}
O.~{Chesneau}, A.~{Meilland}, D.~P.~K. {Banerjee}, J.-B. {Le Bouquin},
  H.~{McAlister}, F.~{Millour}, S.~T. {Ridgway}, A.~{Spang}, T.~{Ten
  Brummelaar}, M.~{Wittkowski}, N.~M. {Ashok}, M.~{Benisty}, J.-P. {Berger},
  T.~{Boyajian}, C.~{Farrington}, P.~J. {Goldfinger}, A.~{Merand},
  N.~{Nardetto}, R.~{Petrov}, T.~{Rivinius}, G.~{Schaefer}, Y.~{Touhami}, and
  G.~{Zins}.
\newblock \emph{\aap}, 534:\penalty0 L11, 2011.

\bibitem[{Colavita} et~al.(1999){Colavita}, {Wallace}, {Hines}, {Gursel},
  {Malbet}, {Palmer}, {Pan}, {Shao}, {Yu}, {Boden}, {Dumont}, {Gubler},
  {Koresko}, {Kulkarni}, {Lane}, {Mobley}, and {van Belle}]{Colavitaetal1999}
M.~M. {Colavita}, J.~K. {Wallace}, B.~E. {Hines}, Y.~{Gursel}, F.~{Malbet},
  D.~L. {Palmer}, X.~P. {Pan}, M.~{Shao}, J.~W. {Yu}, A.~F. {Boden}, P.~J.
  {Dumont}, J.~{Gubler}, C.~D. {Koresko}, S.~R. {Kulkarni}, B.~F. {Lane}, D.~W.
  {Mobley}, and G.~T. {van Belle}.
\newblock \emph{\apj}, 510:\penalty0 505--521, 1999.

\bibitem[{Drake} et~al.(2008){Drake}, {Page}, {Osborne}, {Beardmore}, {Ness},
  {Starrfield}, {Schwarz}, {Tsujimoto}, {Barlow}, {Wesson}, {Bode}, {Corradi},
  {Rodriguez-Gil}, {Drew}, {Gaensicke}, {Steeghs}, {Knigge}, {Takai}, and
  {Zijlstra}]{Drakeetal2008}
J.~J. {Drake}, K.~{Page}, J.~{Osborne}, A.~{Beardmore}, J.-U. {Ness},
  S.~{Starrfield}, G.~{Schwarz}, M.~{Tsujimoto}, M.~{Barlow}, R.~{Wesson},
  M.~{Bode}, R.~{Corradi}, P.~{Rodriguez-Gil}, J.~{Drew}, B.~{Gaensicke},
  D.~{Steeghs}, C.~{Knigge}, D.~{Takai}, and A.~{Zijlstra}.
\newblock \emph{The Astronomer's Telegram}, 1603, 2008.

\bibitem[{Gehrz} et~al.(1998){Gehrz}, {Truran}, {Williams}, {Gehrz}, {Truran},
  {Williams}, and {Starrfield}]{Gehrzetal1998}
R.~D. {Gehrz}, J.~W. {Truran}, R.~E. {Williams}, R.~D. {Gehrz}, J.~W. {Truran},
  R.~E. {Williams}, and S.~{Starrfield}.
\newblock \emph{\pasp}, 110:\penalty0 3--26, 1998.

\bibitem[{Greenstein}(1960)]{McLaughlin1960}
J.~L. {Greenstein}, editor.
\newblock \emph{"{The Spectra of Novae}"}, volume 585, 1960. University of
  Chicago Press, Chicago.

\bibitem[{Hachisu} and {Kato}(2010)]{Hachisu2010}
I.~{Hachisu} and M.~{Kato}.
\newblock \emph{\apj}, 709:\penalty0 680--714, 2010.

\bibitem[{Hernanz} and {Jos{\'e}}(2002)]{Bode2002}
M.~{Hernanz} and J.~{Jos{\'e}}, editors.
\newblock \emph{Classical Nova Explosions}, volume 637 of \emph{American
  Institute of Physics Conference Series}, 2002. AIPC.

\bibitem[{Konacki} et~al.(2009){Konacki}, {Muterspaugh}, {Kulkarni}, and
  {He{\l}miniak}]{Konacki2009}
M.~{Konacki}, M.~W. {Muterspaugh}, S.~R. {Kulkarni}, and K.~G. {He{\l}miniak}.
\newblock {Konacki}, m. and {Muterspaugh}, m.~w. and {Kulkarni}, s.~r. and
  {He{\l}miniak}, k.~g.
\newblock \emph{\apj}, 704:\penalty0 513--521, 2009.

\bibitem[{Lane} et~al.(2005){Lane}, {Retter}, {Thompson}, and {Eisner}]{Lane05}
B.~F. {Lane}, A.~{Retter}, R.~R. {Thompson}, and J.~A. {Eisner}.
\newblock \emph{{\apjl}}, 622:\penalty0 {L137--L140}, 2005.

\bibitem[{Lane} et~al.(2007{\natexlab{a}}){Lane}, {Retter}, {Eisner},
  {Muterspaugh}, {Thompson}, and {Sokoloski}]{Lane07}
B.~F. {Lane}, A.~{Retter}, J.~A. {Eisner}, M.~W. {Muterspaugh}, R.~R.
  {Thompson}, and J.~L. {Sokoloski}.
\newblock \emph{{\apj}}, 669,:\penalty0 {1150--1155}, 2007{\natexlab{a}}.

\bibitem[{Lane} et~al.(2007{\natexlab{b}}){Lane}, {Sokoloski}, {Barry},
  {Traub}, {Retter}, {Muterspaugh}, {Thompson}, {Eisner}, {Serabyn}, and
  {Mennesson}]{Lane07b}
B.~F. {Lane}, J.~L. {Sokoloski}, R.~K. {Barry}, W.~A. {Traub}, A.~{Retter},
  M.~W. {Muterspaugh}, R.~R. {Thompson}, J.~A. {Eisner}, E.~{Serabyn}, and
  B.~{Mennesson}.
\newblock \emph{\apj}, 658:\penalty0 520--524, 2007{\natexlab{b}}.

\bibitem[Lawson(1999)]{Lawson1999}
P.~Lawson.
\newblock \emph{Principles of long baseline interferometry}.
\newblock Wiley-interscience, 1999.

\bibitem[{Lynch} et~al.(2007){Lynch}, {Russell}, {Rudy}, and
  {Woodward}]{Lynch2007}
D.~K. {Lynch}, R.~W. {Russell}, R.~J. {Rudy}, and C.~E. {Woodward}.
\newblock \emph{\iaucirc}, 8883:\penalty0 2, 2007.

\bibitem[{Monnier} et~al.(2006){Monnier}, {Barry}, {Traub}, {Lane}, {Akeson},
  {Ragland}, {Schuller}, {Le Coroller}, {Berger}, {Millan-Gabet}, {Pedretti},
  {Schloerb}, {Koresko}, {Carleton}, {Lacasse}, {Kern}, {Malbet}, {Perraut},
  {Kuchner}, and {Muterspaugh}]{Monnier2006}
J.~D. {Monnier}, R.~K. {Barry}, W.~A. {Traub}, B.~F. {Lane}, R.~L. {Akeson},
  S.~{Ragland}, P.~A. {Schuller}, H.~{Le Coroller}, J.-P. {Berger},
  R.~{Millan-Gabet}, E.~{Pedretti}, F.~P. {Schloerb}, C.~{Koresko}, N.~P.
  {Carleton}, M.~G. {Lacasse}, P.~{Kern}, F.~{Malbet}, K.~{Perraut}, M.~J.
  {Kuchner}, and M.~W. {Muterspaugh}.
\newblock \emph{\apjl}, 647:\penalty0 L127--L130, 2006.

\bibitem[{Nakamura} et~al.(2007){Nakamura}, {Yamaoka}, {Dillon}, {Guido},
  {Sostero}, {Abe}, {Nakano}, and {Labordena}]{Nakamuraetal2007}
Y.~{Nakamura}, H.~{Yamaoka}, W.~G. {Dillon}, E.~{Guido}, G.~{Sostero},
  H.~{Abe}, S.~{Nakano}, and C.~{Labordena}.
\newblock \emph{Central Bureau Electronic Telegrams}, 1029, 2007.

\bibitem[{Nakano} et~al.(2007){Nakano}, {Kadota}, {Waagen}, {Swierczynski},
  {Komorous}, {King}, and {Bortle}]{Nakano2007}
S.~{Nakano}, K.~{Kadota}, E.~{Waagen}, S.~{Swierczynski}, M.~{Komorous},
  R.~{King}, and J.~{Bortle}.
\newblock \emph{{\iaucirc}}, 8861:\penalty0 2--+, 2007.

\bibitem[{Ness} et~al.(2009){Ness}, {Drake}, {Beardmore}, {Boyd}, {Bode},
  {Brady}, {Evans}, {Gaensicke}, {Kitamoto}, {Knigge}, {Miller}, {Osborne},
  {Page}, {Rodriguez-Gil}, {Schwarz}, {Staels}, {Steeghs}, {Takei},
  {Tsujimoto}, {Wesson}, and {Zijlstra}]{Ness2009}
J.-U. {Ness}, J.~J. {Drake}, A.~P. {Beardmore}, D.~{Boyd}, M.~F. {Bode},
  S.~{Brady}, P.~A. {Evans}, B.~T. {Gaensicke}, S.~{Kitamoto}, C.~{Knigge},
  I.~{Miller}, J.~P. {Osborne}, K.~L. {Page}, P.~{Rodriguez-Gil}, G.~{Schwarz},
  B.~{Staels}, D.~{Steeghs}, D.~{Takei}, M.~{Tsujimoto}, R.~{Wesson}, and
  A.~{Zijlstra}.
\newblock \emph{\aj}, 137:\penalty0 4160--4168, 2009.

\bibitem[Payne-Gaposchkin(1957)]{Payne1957}
C.~Payne-Gaposchkin.
\newblock \emph{The Galactic Novae}.
\newblock North Holland Publishing Co., Amsterdam, 1957.

\bibitem[{Poggiani}(2008)]{Poggiani2008}
R.~{Poggiani}.
\newblock \emph{\apss}, 315:\penalty0 79--85, 2008.

\bibitem[{Prater} et~al.(2007){Prater}, {Rudy}, {Lynch}, {Mazuk}, {Perry}, and
  {Puetter}]{Prater2007}
T.~R. {Prater}, R.~J. {Rudy}, D.~K. {Lynch}, S.~{Mazuk}, R.~B. {Perry}, and
  R.~C. {Puetter}.
\newblock \emph{\iaucirc}, 8904:\penalty0 2, 2007.

\bibitem[{Prialnik} and {Kovetz}(1995)]{Prialniketal1995}
D.~{Prialnik} and A.~{Kovetz}.
\newblock \emph{\apj}, 445:\penalty0 789--810, 1995.

\bibitem[{Retter}(2003)]{Retter2003}
A.~{Retter}.
\newblock "{The Transition Phase in Classical Novae}".
\newblock In "{R.~L.~M.~Corradi, J.~Mikolajewska, \& T.~J.~Mahoney}", editor,
  \emph{Astronomical Society of the Pacific Conference Series}, volume 303 of
  \emph{Astronomical Society of the Pacific Conference Series}, page 232, 2003.

\bibitem[{Rodr{\'{\i}}guez-Gil} et~al.(2010){Rodr{\'{\i}}guez-Gil},
  {Santander-Garc{\'{\i}}a}, {Knigge}, {Corradi}, {G{\"a}nsicke}, {Barlow},
  {Drake}, {Drew}, {Miszalski}, {Napiwotzki}, {Steeghs}, {Wesson}, {Zijlstra},
  {Jones}, {Liimets}, {Mu{\~n}oz-Darias}, {Pyrzas}, and
  {Rubio-D{\'{\i}}ez}]{Rodriguez2010}
P.~{Rodr{\'{\i}}guez-Gil}, M.~{Santander-Garc{\'{\i}}a}, C.~{Knigge}, R.~L.~M.
  {Corradi}, B.~T. {G{\"a}nsicke}, M.~J. {Barlow}, J.~J. {Drake}, J.~{Drew},
  B.~{Miszalski}, R.~{Napiwotzki}, D.~{Steeghs}, R.~{Wesson}, A.~A. {Zijlstra},
  D.~{Jones}, T.~{Liimets}, T.~{Mu{\~n}oz-Darias}, S.~{Pyrzas}, and M.~M.
  {Rubio-D{\'{\i}}ez}.
\newblock \emph{\mnras}, 407:\penalty0 L21--L25, 2010.

\bibitem[{Samus}(2007)]{Samus2007}
N.~N. {Samus}.
\newblock \emph{\iaucirc}, 8863:\penalty0 2, 2007.

\bibitem[{Shaviv}(2001)]{Shaviv2001}
N.~J. {Shaviv}.
\newblock \emph{\mnras}, 326:\penalty0 126--146, 2001.

\bibitem[{Shaviv}(2002)]{Shaviv2002}
N.~J. {Shaviv}.
\newblock "{Classical Novae as Super-Eddington Objects}".
\newblock In "{M.~Hernanz \& J.~Jos{\'e}}", editor, \emph{Classical Nova
  Explosions}, volume 637 of \emph{American Institute of Physics Conference
  Series}, pages 259--265, 2002.

\bibitem[{Starrfield} and {Sparks}(1987)]{Starrfield1987}
S.~{Starrfield} and W.~M. {Sparks}.
\newblock \emph{\apss}, 131:\penalty0 379--393, 1987.

\bibitem[{Tanaka} et~al.(2011){Tanaka}, {Nogami}, {Fujii}, {Ayani}, and
  {Kato}]{Tanaka2011}
J.~{Tanaka}, D.~{Nogami}, M.~{Fujii}, K.~{Ayani}, and T.~{Kato}.
\newblock \emph{\pasj}, 63:\penalty0 159--, 2011.

\bibitem[{Tarasova}(2007)]{Tarasova2007}
T.~N. {Tarasova}.
\newblock \emph{Information Bulletin on Variable Stars}, 5807, 2007.

\bibitem[{Tsujimoto} et~al.(2009){Tsujimoto}, {Takei}, {Drake}, {Ness}, and
  {Kitamoto}]{Tsujimoto2009}
M.~{Tsujimoto}, D.~{Takei}, J.~J. {Drake}, J.-U. {Ness}, and S.~{Kitamoto}.
\newblock \emph{\pasj}, 61:\penalty0 69, 2009.

\bibitem[{Vogt}(1987)]{Vogt1987}
S.~S. {Vogt}.
\newblock \emph{\pasp}, 99:\penalty0 1214--1228, 1987.

\bibitem[{Wade} et~al.(2000){Wade}, {Harlow}, and {Ciardullo}]{Wade2000}
R.~A. {Wade}, J.~J.~B. {Harlow}, and R.~{Ciardullo}.
\newblock \emph{\pasp}, 112:\penalty0 614--624, 2000.

\bibitem[{{Wade}, R.~A. and {Hubeny}, I.}(1998)]{Wade1998}
{{Wade}, R.~A. and {Hubeny}, I.}
\newblock \emph{{\apj}}, 509:\penalty0 {350--361}, 1998.

\bibitem[{Warner}(1995)]{Warner1995}
B.~{Warner}.
\newblock "{Cataclysmic variable stars.}".
\newblock \emph{Cambridge Astrophysics Series}, 28, 1995.

\bibitem[{Warner}(2008)]{Warner2008}
B.~{Warner}.
\newblock \emph{Classical Novae}, volume 2nd Edition.
\newblock Cambridge: Cambridge University Press, 2008.

\bibitem[{Wesson} et~al.(2008){Wesson}, {Barlow}, {Corradi}, {Drew}, {Groot},
  {Knigge}, {Steeghs}, {Gaensicke}, {Napiwotzki}, {Rodriguez-Gil}, {Zijlstra},
  {Bode}, {Drake}, {Frew}, {Gonzalez-Solares}, {Greimel}, {Irwin},
  {Morales-Rueda}, {Nelemans}, {Parker}, {Sale}, {Sokoloski}, {Somero},
  {Uthas}, {Walton}, {Warner}, {Watson}, and {Wright}]{Wessonetal2008}
R.~{Wesson}, M.~J. {Barlow}, R.~L.~M. {Corradi}, J.~E. {Drew}, P.~J. {Groot},
  C.~{Knigge}, D.~{Steeghs}, B.~T. {Gaensicke}, R.~{Napiwotzki},
  P.~{Rodriguez-Gil}, A.~A. {Zijlstra}, M.~F. {Bode}, J.~J. {Drake}, D.~J.
  {Frew}, E.~A. {Gonzalez-Solares}, R.~{Greimel}, M.~J. {Irwin},
  L.~{Morales-Rueda}, G.~{Nelemans}, Q.~A. {Parker}, S.~E. {Sale}, J.~L.
  {Sokoloski}, A.~{Somero}, H.~{Uthas}, N.~A. {Walton}, B.~{Warner}, C.~A.
  {Watson}, and N.~J. {Wright}.
\newblock \emph{\apjl}, 688:\penalty0 L21--L24, 2008.

\bibitem[{{Williams}, R.~E.}(1992)]{Williams1992}
{{Williams}, R.~E.}
\newblock \emph{{\aj}}, 104,:\penalty0 {725--733}, 1992.

\end{thebibliography}
\bibliographystyle{plainnat}

\end{document}